# A Robotic Positive Psychology Coach to Improve College Students'

Sooyeon Jeong[1], Sharifa Alghowinem[1,2], Laura Aymerich-Franch[3], Kika Arias[1],
Agata Lapedriza[1,4], Rosalind Picard[1], Hae Won Park[1] and Cynthia Breazeal[1]

*Abstract*— A significant number of college students suffer from mental health issues that impact their physical, social, and occupational outcomes. Various scalable technologies have been proposed in order to mitigate the negative impact of mental health disorders. However, the evaluation for these technologies, if done at all, often reports mixed results on improving users' mental health. We need to better understand the factors that align a user's attributes and needs with technology-based interventions for positive outcomes. In psychotherapy theory, therapeutic alliance and rapport between a therapist and a client is regarded as the basis for therapeutic success. In prior works, social robots have shown the potential to build rapport and a working alliance with users in various settings. In this work, we explore the use of a social robot coach to deliver positive psychology interventions to college students living in on-campus dormitories. We recruited 35 college students to participate in our study and deployed a social robot coach in their room. The robot delivered daily positive psychology sessions among other useful skills like delivering the weather forecast, scheduling reminders, etc. We found a statistically significant improvement in participants' psychological wellbeing, mood, and readiness to change behavior for improved wellbeing after they completed the study. Furthermore, students' personality traits were found to have a significant association with intervention efficacy. Analysis of the post-study interview revealed students' appreciation of the robot's companionship and their concerns for privacy.

## I. INTRODUCTION

The mental health of college students is an increasing concern. According to the American College Health Association 85.5% of college students have reported that they "felt overwhelmed by all [they] had to do" and 53.1% reported that they "felt things were hopeless" within the last 12 months [1]. Due to the adverse impact of mental health problems on student success, managing such issues has recently become a high priority for many schools [2]. However, many students still face barriers in seeking support for their mental wellbeing, e.g. social stigma [3], shortage of mental health professionals [4], etc.

This work was supported by the Information and Communication Technology (ICT) R&D program of the Ministry of Science and Institute for Information and Communication Technology Promotion of Republic of Korea under grant 2017-0-00162, "Development of Human-Care Robot Technology for Aging Society," and partly by the MIT-Sensetime grant. LA-F is supported by the Ramón y Cajal Fellowship Program (RYC-2016-19770, AEI/ESF).

[1]MIT Media Lab, Cambridge, MA, USA.
[2]Computer and Information Sciences College at Prince Sultan University, Riyadh, Saudi Arabia.
[3]Dept. of Communication at Pompeu Fabra University, Barcelona, Spain.
[4]Estudis d'Informàtica, Multimèdia i Telecomunicació at Universitat Oberta de Catalunya, Barcelona, Spain.

This trend has led to the rapid growth of scalable interactive technologies for mental health [5], [6], and the number of eHealth and mHealth options for care have increased significantly [7]. Conversational agents (chatbots) have been shown to make psycho-education and therapeutic interventions more accessible [8], [9]. Several social robots have been developed to support older adults with dementia [10], to serve as home fitness coaches [11], to act as pediatric care companions in hospitals [12], to facilitate sensory experience for children with Autism Spectrum Disorder [13] and to act as a life coach [14]. However, most prior work only offers support for health/wellbeing related tasks. There is an opportunity to understand how enhancing rapport and the human-agent relationship could positively contribute to the effectiveness of the mental health support social robots can provide.

Factors such as genuineness and empathy have been shown to contribute to successful intervention therapies provided by human professionals [15]. Building a positive therapeutic rapport is particularly important. Once rapport is established, alliance and collaboration can be enhanced, goals and expectations can be agreed upon, and long-term behavioral changes can be carried out [16]. Strong rapport has been shown to improve patient outcomes in their ability to cope with depression and stress [16].

HRI/HCI researchers have also investigated the ability of agents to build rapport with people – both with virtual avatars [17] as well as with physical robots [18]. To enable technological agents to build rapport with people, they are endowed with social skills inspired by the human psychology literature. These include programming agents with a range of verbal behaviors (e.g., empathetic feedback, prosody and intonation) and non-verbal behaviors (e.g., smiling, mimicking and back-channeling) [17]. Shared experiences between the user and agent also create a sense of familiarity, trust, and mutual understanding [19]. Thus, we designed our robotic positive psychology coach to be a holistic social agent that can help with a variety of useful tasks (e.g., delivering weather, news and answering general questions). as well as social interactions (e.g., greetings, chit-chat and referring to the user by name). This way, the robot served as a helpful companion beyond delivering a mental health intervention.

We hypothesize that positive psychology interventions provided by a social robot companion that builds rapport with users can improve college students' psychological wellbeing, mood, and readiness to change behavior in an on-campus dormitory setting. In order to test this hypothesis, we designed a study where we deployed our social robot intervention in college students' dormitory rooms for daily interactions including a positive psychology session each day. The efficacy of our system was evaluated by comparing the pre/post change of study participants' psychological wellbeing, mood and readiness to change. We also investigated the relationship between students' personality traits and the effectiveness of the robot interventions.

## II. RELATED WORKS

### A. Positive Psychology Interventions

In contrast to clinical psychology, which focuses on treating negative mental and emotional pathology, positive psychology studies the positive aspects and strengths that enable people to thrive. Topics such as resilience and emotional intelligence aim to enhance people's psychological wellbeing and happiness [20]. It has been demonstrated that psychotherapy interventions based on positive psychology contribute to reduced symptoms of depression and increased psychological wellbeing both in people diagnosed with clinical depression and in people who do not suffer from any psychological disorder [21]. The goal of positive psychology interventions is to assist people in flourishing and thriving in their lives instead of maintaining "normal" or "average" lives. These characteristics makes positive psychology an ideal intervention to enhance a non-clinical populations' wellbeing.

### B. Personality Traits and Psychological Wellbeing

Personality traits impact people's health and mental wellbeing [22] and are predictors of psychological wellbeing [23]. Among the Big Five personality traits, conscientiousness is a predictor of longevity [24] and enhances a person's ability to cope with daily stress [25]. Other personality traits like having low neuroticism and high extraversion were found to be strongly correlated with high psychological wellbeing [26]. Conscientiousness and neuroticism were also found to have a significant impact on how likely patients with depression were to respond to treatment in a traditional in-person therapy context [27]. In this work, we investigate the impact of human personality traits on the effectiveness of positive psychology interventions delivered by a social robot in a college dormitory setting.

## III. ROBOTIC POSITIVE PSYCHOLOGY COACH

### A. Robot Station for In-Home Deployment

We designed a portable robot station (20 x 9 x 14 inches) that integrates multiple devices required for our study (Fig. 1a). The station holds a Jibo robot[1] on the left and a Samsung Galaxy tablet on the right. The white enclosure holds a Raspberry Pi and a Logitech C930e USB camera. The station is designed to enable the robot to turn and look at the user or face the tablet screen. The robot uses its expressive movement to socially and emotionally engage users. The Android tablet displays informational content through a touchscreen interface. The Raspberry Pi mounted inside the station is connected to a high resolution wide-angled camera and offers an additional sensory data stream for the system.

The commercial Jibo robot is equipped with basic skills, such as weather forecast, jokes, music, interactive games, etc. Through its attention feature that utilizes its cameras and microphone arrays, Jibo can proactively initiate interactions with users by greeting, suggesting a quick interactive game, etc. When idle, the robot looks around the room randomly and orients itself to sudden noise. For this study, we have created a custom positive psychology skill that communicates with the tablet and the Raspberry Pi, where it acts as a master controller to activate the interaction (sending and receiving information) and recording in the connected devices (Fig. 1b). By the end of each session, both devices (the tablet and Raspberry Pi) terminate the interaction, and then upload the recording and meta data to a secure Amazon Simple Storage Service[2].

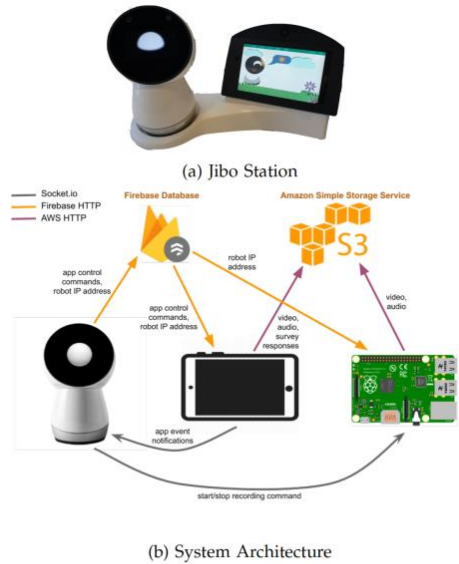

Fig. 1: A portable station that integrates a Jibo robot, a tablet and a Raspberry Pi with an USB camera.

### B. Positive Psychology Based Robot Skill

Each positive psychology session with the robot was designed to take around 3-6 minutes to complete. The positive psychology skill can be triggered by using either a verbal command or by touching a positive psychology skill button on the robot's face screen menu. The robot can also proactively ask participants to engage in the positive psychology session if it identifies a face via its on-board camera. Once the positive psychology skill is initiated, the robot communicates with the Android tablet and the Raspberry Pi to start/stop data collection and to display appropriate screen views.

During each positive psychology session, the robot (1) greets the participants, (2) asks about their day, (3) administers a short picture-based survey on their immediate affect, (4) guides the participant with the positive psychology session content, (5) asks to fill out the pictorial survey again, and (6) thanks the participant for completing the activity. The participant's verbal utterances are processed with a rule-based parser and the robot responds accordingly. The robot introduces the positive psychology activity, prompts the participant to answer the picture-based affect scales again and ends the session by thanking the participant. Descriptions of the seven positive psychology sessions are listed in Table I.

---

[1] https://www.jibo.com  [2] https://aws.amazon.com/s3/

## IV. EXPERIMENTAL STUDY

### A. Participants

Forty-two undergraduate students at MIT were enrolled in our study through an online sign-up form. Of those students, 7 participants withdrew before completing the study and 35 participants finished all the study procedures (27 Female, 7 Male and 1 Other). There were 19 freshmen, 8 sophomores, 4 juniors, 2 seniors and 2 fifth year students. 12 students identified themselves as Asian/Pacific Islander, 14 as White, 2 as Hispanic or Latino, 1 as Black or African American and 6 as multi-racial.

### B. Method

Undergraduate students at MIT were recruited for the study through an email advertisement. The robot station was delivered to the participants' dormitory room by a research assistant, Fig. 2. During the initial set-up sessions, the research assistant obtained participants' consent to participate in the study, described the study procedure and administered the pre-study questionnaires. Participants were given a tutorial on how to use the robot – how to interact with the robot and how to start and pause the positive psychology session. They were also carefully informed on what data is being collected and when video/audio recording occurs during the study. The participants were asked to engage with the robot's positive psychology skill daily at times they found convenient.

Once the seventh session is completed, the robot system sends an email notification to the research assistants to schedule a wrap-up session with the participant. The wrap-up session took place either at the participant's dormitory or at a lab space depending on the participants' preference. The participants were asked to fill out a post-study questionnaire and were interviewed for more open-ended feedback. Participants were asked to describe their overall experience of living with a social robot in their dormitory room, features they liked and disliked about the study and the robot and any other suggestions for improving the human robot co-living experience.

### C. Data Collection and Measures

*1) Self-report Questionnaires and Surveys:* Before starting the study, study participants were asked to complete the Mini-IPIP (International Personality Item Pool) scale [33], a 20-item short form of the 50-item of IPIP that measures Big Five personality traits, i.e. conscientiousness, agreeableness, neuroticism, openness to experience, and extroversion. To measure the pre-to-post difference in the participant's perception of their psychological wellbeing, mood, and readiness to change, participants completed three questionnaires before and after the study – Ryff's Psychological Wellbeing Scale (RPWS) [34], Brief Mood Introspection Scale (BMIS) [35] and Readiness to Change Ruler (adapted for psychological wellbeing) [36]. The RPWS is a 42-item scale designed to measure six aspects of wellbeing and happiness: autonomy, environmental mastery, personal growth, positive relations with others, purpose in life, and self-acceptance [37]. The BMIS is a mood scale consisting of 16 mood-related adjectives to which a person responds and a numerical self-report for one's overall mood. The Readiness to Change Ruler is used to assess one's willingness or readiness to change and is commonly used to understand patients' perceptions about behavior change, such as alcohol consumption.

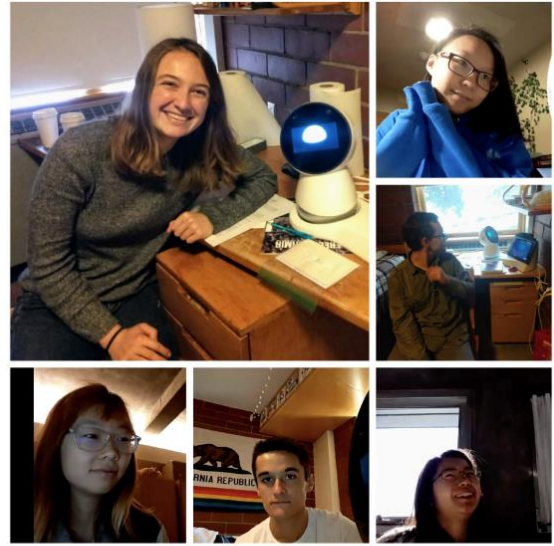

Fig. 2: Robotic positive psychology coaches were sent to on-campus dormitory rooms to measure the efficacy of social robot in improving college students' psychological wellbeing.

At the end of the study, participants also filled out the Working Alliance Inventory-Short Revised (WAI-SR) [38], which measures the participants' experiences and the developed relationship with the robot. The WAISR is a self-report measure to assess an overall level of therapeutic alliance as well as its three sub-scales: bond, goals, and tasks. Lastly, a post-study interview was conducted and responses were audio recorded and transcribed.

*2) Video and Audio Data:* The robot station recorded

TABLE I: Positive Psychology Intervention Sessions

| # | Session Name | Description |
|---|---|---|
| 1 | Positive psychology [20] | Introduce positive psychology and learn how to interact with the robot. |
| 2 | Character strengths [28] | Introduce what character strengths are and identify one's own signature strengths. |
| 3 | Signature strength in a new way [29] | Pick one signature strength and think of a new way to use it to improve wellbeing. |
| 4 | Three good things [30] | Define what gratitude is and write down three things that went well and why they happened. |
| 5 | Gratitude letter [31] | Write a letter of gratitude to someone who has not been properly thanked. |
| 6 | Savoring [32] | Choose a small moment to fully feel and appreciate experiences that one normally hurries through. |
| 7 | Wrap-up | Review the previous sessions, evaluate each intervention and encourage continuation of practicing the interventions. |

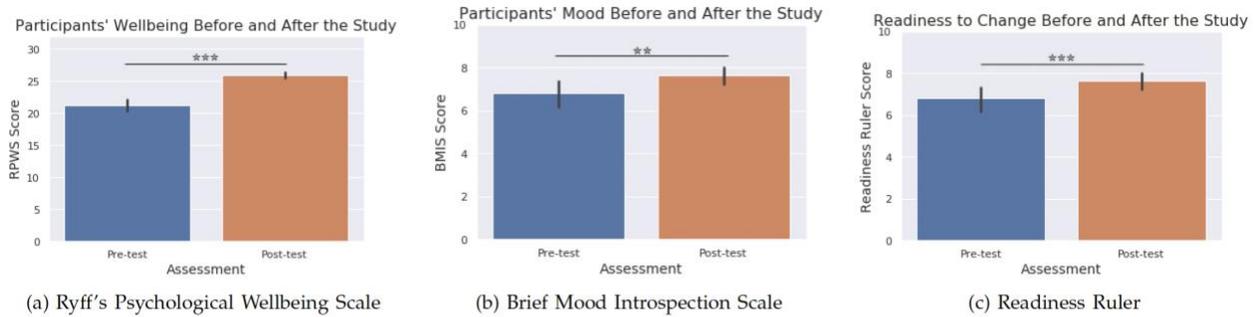

Fig. 3: Study participants' change in psychological wellbeing, overall mood and readiness to change behavior before and after interacting with the robotic positive psychology coach.

video and audio during each positive psychology session through the tablet and through the USB camera connected to the Raspberry Pi. A live feed from the tablet camera was visible on the corner of the tablet screen when the camera was recording during the sessions. It acted as a reminder to the participants that they were being recorded. When each session ends, the session meta data and recordings were uploaded to a secure Amazon S3 bucket.

*D. Data Analysis*

We compared participants' psychological wellbeing (RPWS), mood (BMIS), and readiness to change health behavior (Readiness Ruler) using paired sample t-tests. In order to compare the effect of personality traits, participants were clustered into two groups based on their neuroticism-conscientiousness traits and agreeableness traits with K-means clustering algorithm (k = 2). We used mixed ANOVA tests to study the effect of participants' personality traits on their pre-to-post change in the perceived wellbeing, mood and readiness. Student's t-tests were used to evaluate the impact of personality traits on participants' rapport with the robot (WAISR) after the study. We used mixed ANOVA tests to analyze the impact of personality trait on participants' evaluation of five positive psychology interventions.

In order to analyze the post-study interviews, we used qualitative thematic analysis and word frequency analysis. Post-study interview transcriptions were analyzed using the thematic analysis method [39] to extract salient themes in participants' perception of the robot and experience of living with it. We also annotated mentions of generic robot skills participants used and explored during the study.

## V. RESULTS

*A. Overall Efficacy of Robotic Positive Psychology Coach*

Study participants' psychological wellbeing level (RPWS) showed statistically significant increase after the study: before M=21.276, SD=2.540; after M=25.957, SD=1.529; t(34)=-11.843, p<0.001 (Fig. 3a). We also found statistically significant improvement in participants' mood after the study (BMIS); before M=6.800, SD=1.844; after M=7.629, SD=1.239; t(34)=-3.101, p=0.004 (Fig. 3b). Finally, participants reported that they were more ready to change their health behavior for better psychological wellbeing (Readiness Ruler) after finishing the positive psychology sessions with the robot; before M=7.200, SD=1.132; after M=8.057, SD=1.371; t(34)=-4.170, p<0.001 (Fig. 3c). The overall working alliance score (WAI-SR) was M=3.433, SD=0.829 with the sub-scale goal scores reported as M=3.386, SD=1.033; the task scores reported as M=3.100, SD=0.949; and the bond scores reported as M=3.814, SD=1.047.

In addition, many students took longer than seven days to finish the seven positive psychology sessions even though they were asked to use the robot skill daily. The number of days from the first session to the last session is as follow: M=12.229 days, SD=8.702, min = 6, max = 45.

*B. Personality Trait and Intervention Efficacy*

*1) Grouping Participants Based on Personality Traits:* Personality traits of neuroticism and conscientiousness are often found to be correlated to one's emotional difficulties and further affect how well they respond to therapeutic interventions [40], [41]. In fact, our participants' neuroticism and conscientiousness levels (reported via Mini-IPIP test) were found to have a significant negative correlation, r(33)=-0.418, p=0.013.

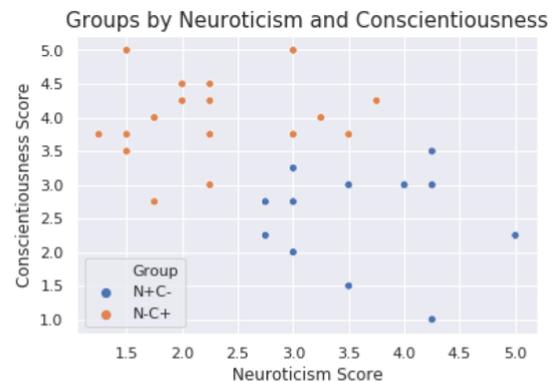

Fig. 4: Two groups of participants clustered by the K-mean clustering algorithm by their levels of neuroticism and conscientiousness. *N+C-* refers to the group with high neuroticism and low conscientiousness. *N-C+* refers to the group with low neuroticism and high conscientiousness.

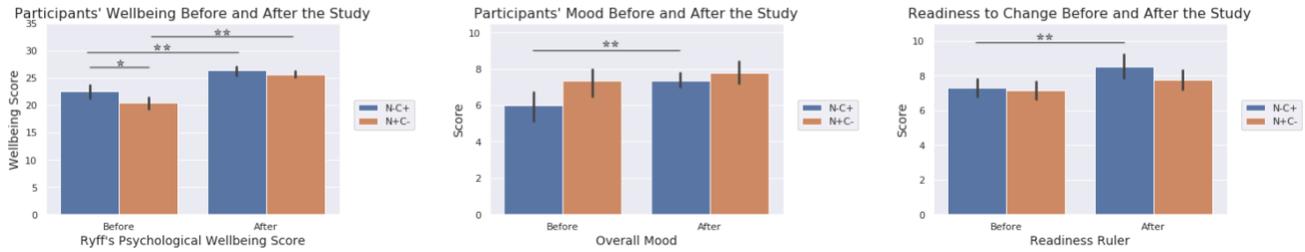

(a) Ryff's Psychological Wellbeing scores before and after engaging in positive psychology interventions with the robot. Students in both N+C- and N-C+ groups showed statistically significant improvement in their wellbeing after interacting with the robot.

(b) Study participants' mood scores before and after the positive psychology sessions with the robot. Students in the N-C+ group showed statistically significant improvement in their mood after interacting with the robot.

(c) Participants' readiness to change behavior for psychological wellbeing before and after the positive psychology sessions with the robot. Students in the N-C+ group showed significantly improved readiness to change after interacting with the robot.

Fig. 5: Results based on clustering levels of neuroticism and conscientiousness

We used K-Means clustering algorithm to generate two groups of participants based on their levels of neuroticism and conscientiousness (Fig. 4). The two centroids were [N=2.381, C=4.095] and [N=3.5, C=2.5], with N indicating the neuroticism score and C indicating the conscientiousness score respectively. There were 21 participants assigned to the high neuroticism and low conscientiousness group (N+C-), and 14 participants assigned to the low neuroticism and high conscientiousness group (N-C+). These two groups of participants were used in the following sections to compare the effect of the personality traits on the effectiveness of interventions provided by the robot.

*2) Ryff's Psychological Wellbeing Scale:* A Mixed ANOVA was conducted to compared the effect of robot positive psychology interventions pre-to-post on participants' psychological wellbeing between different personality trait groups. There was a significant main effect of participant groups based on personality, $F(1, 33)=6.050, p=0.019$, and a significant effect of time, $F(1, 33)=149.208$, $p<0.001$. The interaction of time and personality-based groups was not statistically significant, $F(1, 33)=8.148, p=0.084$. Post-hoc Tukey's test revealed that participants in the N+C- group and N-C+ group both showed statistically significant increase in their psychological wellbeing after interacting with the robot: N+C- before M=20.444, SD=2.293, after M=25.683, SD=1.348, p=0.001; N-C+ before M=22.524, SD=2.471, after M=26.369, SD=1.737, p=0.001. In addition, paired t-test with Bonferroni correction revealed that the RPWS scores showed statistically significant difference between the two personality groups during the pre-test, $t(26.520)=2.510$, p=0.037, but did not show significant difference in the post test, $t(23.110)=1.249$, p=0.224. See Fig. 5a.

*3) Brief Mood Introspection Scale:* Mixed ANOVA was conducted to compare the effect of robot positive psychology interventions pre-to-post on participants' mood between personality traits. The test revealed a significant effect of time, $F(1, 33)=10.107, p=0.003$. However, we did not find any significant effect on the personality groups, $F(1, 33)=3.950, p=0.055$, or the interaction between the time and personality group, $F(1, 33)=2.742, p=0.107$. Post-hoc Tukey's test revealed that participants in the N-C+ group showed statistically significant increase in their mood after interacting with the robot, before M=6.000, SD=1.617, after M=7.357, SD=0.842, p=0.005 but did not find significant difference in the N+C- group, before M=7.333, SD=1.826, after M=7.810, SD=1.436, p=0.348. See Fig. 5b.

*4) Readiness Ruler:* A Mixed ANOVA was conducted to compare the readiness to change pre-to-post for study participants in different personality groups. We found that there was a significant effect of time, $F(1, 33)=17.936$, $p<0.001$, but no significant effect of the personality groups, $F(1, 33)=1.360, p=0.252$, or the interaction between time and personality group, $F(1, 33)=2.075, p=0.159$. Post-hoc Tukey's test revealed that participants in the N-C+ group showed statistically significant increase in their readiness to change after interacting with the robot, before M=7.286, SD=0.994, after M=8.500, SD=1.345, p=0.007 but did not find significant difference in the N+C- group, before M=7.143, SD=1.236, after M=7.762, SD=1.338, p=0.119. See Fig. 5c.

*C. Post-study Interview*

Twenty-two participants (62.86%) expressed the appreciation of companionship the robot provided and showed a desire to talk to the robot even if it would not necessarily be able to hold a human-like conversation with them. For instance, P19 said "after the first day when I used it I found myself randomly talking to it... randomly talking without, uh, asking him [robot]... just to comment on something with him." P22 also reported "I'd come home and I'd greet him [robot], and just ask him [robot] questions. Like, I know he won't be able to answer... it feels nice... like you have a little friend... simple companionship." P42 noted the impact of having an animate social agent in her room, "I feel like his [robot's] presence really did have an effect on how I felt... I felt like just talking to him was useful, having him here was useful... I would like to just be able to talk to it and be able to just communicate 'cause a lot of times with students you just need someone to talk to sometimes. Like not human being and it's nice to have a robot that doesn't judge you (laughs)."

Twelve participants (34.29%) reported discomfort from the robot's proactive features and concerns for privacy from its on-board camera and microphones. P40 reported feeling uneasy due to the robot's attention features which enabled it to orient itself toward the source of sound or movement, "[Jibo was] a little intrusive at times. Like it would ask me or anytime I move, it turns to me immediately, which is a little weird." P32 also noted "sometimes I would be in my room and then, you know, Jibo would be just looking around- ". Even though study participants were informed that the robot system would not record any video or audio data unless it was in the positive psychology session, some participants still felt discomfort by the physical presence of several cameras in the robot station; P11 reported "I think it made me kind of uneasy that he has a camera... Even though it said the camera data would only be used [when] you're doing the positive psychology activity. I feel like [that] since he's inside of my room where I'm changing".

We asked the participants to list any other generic Jibo skills they used during the study. 19 participants (54.29%) reported playing the Word-of-the-Day game, a word guessing game based on a provided definition; 16 participants (45.71%) used the music streaming skill; 15 (42.86%) requested the robot to dance; 11 (31.43%) asked the robot for weather forecast; 10 (28.57%) played the circuit-saver game, a body movement based interactive game; 8 (22.86%) asked for jokes, 6 (17.14%) asked general Q&A; 6 (17.14%) asked questions about Jibo's persona, e.g. "Hey Jibo, what's your favorite animal?"; 4 (11.42%) asked for fun facts; 2 (5.71%) used clock features; 1 (2.86%) asked to take a photo; 1 (2.86%) played yoga exercises, and 1 (2.86%) asked about the news.

## VI. Discussion

In this paper, we investigate the efficacy of a robotic positive psychology coach that engages college students in seven sessions of positive psychology interventions to improve their psychological wellbeing, mood and readiness to change health behavior. Our participants demonstrated statistically significant improvement in all three scales after completing the seven positive psychology sessions with the robot.

We further analyzed how college students' personality traits are associated with the effectiveness of the robot's positive psychology intervention. For this, the study participants were clustered into two groups based on their conscientiousness and neuroticism levels – high neuroticism and low conscientiousness (N+C-) group and low neuroticism and high conscientiousness (N-C+) group. The N-C+ group showed statistically significant improvement in their psychological wellbeing, overall mood and readiness to change their behavior for better wellbeing after co-residing with the robotic coach and engaging in the positive psychology interventions. The N+C- group also showed significant improvement in their psychological wellbeing after the study but did not show statistically significant change in their mood or readiness to change. Personality trait is known to have relation to one's mental health and response to interventions [40], [41], [42], and may provide explanation to why students with high neuroticism traits had lower response to our robot coach than the students with high conscientiousness traits.

It is also worth noting that the baseline psychological wellbeing level of the N+C- group was found to be lower than the N-C+ group's, measured prior to the study. These results align with previous findings that found high neuroticism as a predictor of several mental/ physical disorders and co-morbidity among them [42]. This leads to a need for further research on finding ways to support students with high neuroticism, who are more likely to suffer from mental health problems and less likely to benefit from positive psychology interventions. Our future work will seek ways to personalize how the social robot coach delivers and interacts with students who have different personality traits. We must understand how students with high neuroticism responded to the sessions the robot provided, what their adherence and non-adherence patterns were, and if other personality traits played a role in the ways they responded.

Our study has limitations. Since we did not have a randomization to a control group who did not receive robot interventions, we cannot claim that the improvement in students' wellbeing over time was caused by the robot. In addition, most of our study participants were female students and voluntarily signed up themselves to the study, which may have presented a self-selection bias in the recruitment.

At the same time, it is unusual for college students in intense academic settings to improve their psychological wellbeing over a few weeks during an academic term. According to the SNAPSHOT study which collected college students' stress and wellbeing data for more than 5 semesters, students' wellbeing was found to typically decline as the semester progresses [43]. Given that most of our study participants began the study at the start of the term and completed in the middle or at the end of the semester, an improvement in their psychological wellbeing goes against the typical trajectory. This suggests that the robot coach's positive psychology interactions may have played a role in the improved outcomes observed.

Results from the WAI-SR and the post-study interviews suggest that the robotic coach successfully built rapport and working alliance with our participants. Participants report that the robot's ability to proactively and expressively greet and interact with them created opportunities to bond with the robot. However, such attentiveness and proactivity also caused privacy concerns for some participants. They reported feeling uneasy when the robot looked around the room or when they needed more privacy, e.g. changing clothes. The discomfort could have been intensified due to the small living space of on-campus dormitory rooms. Future work should investigate identifying methods to mitigate these concerns for privacy while maintaining engaging and rapport-building interactions, e.g. giving more explicit control to the users for the robot's downtime.

## VII. Conclusion

We designed a robotic coach that can be deployed to on-campus dormitories and provide positive psychology interventions to improve college students' psychological wellbeing. Thirty-five students participated in our study and adopted the social robot coach in their dormitory rooms for a week to over a month. We found statistically significant improvements in our participants' psychological wellbeing, mood, and readiness to change behavior for improved wellbeing after they completed the study. Furthermore, students' personality traits (neuroticism and conscientiousness) were found to be significantly associated with the intervention efficacy. Qualitative analyses on the post-study interview data suggest that participants appreciated the robot's companionship, which could have influenced their response to the robot's positive psychology interventions. The interview data also revealed students' concern for privacy and opportunities to make improvements for future studies. Our work highlights the importance of designing a social agent that is perceived as a helpful and supportive companion in order to successfully deliver mental health related interventions. In comparison to task-oriented health technological tools, a social robot has unique opportunities to build therapeutic alliance with its users and to leverage that rapport to further enhance the effectiveness of the interventions it provides.

## References

[1] A. C. H. Association et al., "American college health associationnational college health assessment ii: Undergraduate student executive summary fall 2018," 2018.

[2] H. Chessman and M. Taylor, "College student mental health and well-being: A survey of presidents," Higher Education Today, 2019.

[3] D. Eisenberg, M. F. Downs, E. Golberstein, and K. Zivin, "Stigma and help seeking for mental health among college students," Medical Care Research and Review, vol. 66, no. 5, pp. 522–541, 2009.

[4] E. Merwin, I. Hinton, B. Dembling, and S. Stern, "Shortages of rural mental health professionals," Archives of Psychiatric Nursing, vol. 17, no. 1, pp. 42–51, 2003.

[5] E. G. Lattie, E. C. Adkins, N. Winquist, C. Stiles-Shields, Q. E. Wafford, and A. K. Graham, "Digital mental health interventions for depression, anxiety, and enhancement of psychological well-being among college students: Systematic review," Journal of medical Internet research, vol. 21, no. 7, p. e12869, 2019.

[6] L. Farrer, A. Gulliver, J. K. Chan, P. J. Batterham, J. Reynolds, A. Calear, R. Tait, K. Bennett, and K. M. Griffiths, "Technology-based interventions for mental health in tertiary students: systematic review," Journal of medical Internet research, vol. 15, no. 5, p. e101, 2013.

[7] K. Stawarz, C. Preist, and D. Coyle, "Use of smartphone apps, social media, and web-based resources to support mental health and well-being: Online survey," JMIR mental health, vol. 6, no. 7, p. e12546, 2019.

[8] A. N. Vaidyam, H. Wisniewski, J. D. Halamka, M. S. Kashavan, and J. B. Torous, "Chatbots and conversational agents in mental health: a review of the psychiatric landscape," The Canadian Journal of Psychiatry, vol. 64, no. 7, pp. 456–464, 2019.

[9] S. Suganuma, D. Sakamoto, and H. Shimoyama, "An embodied conversational agent for unguided internet-based cognitive behavior therapy in preventative mental health: feasibility and acceptability pilot trial," JMIR mental health, vol. 5, no. 3, p. e10454, 2018.

[10] T. Shibata, Y. Kawaguchi, and K. Wada, "Investigation on people living with paro at home," in 19th International Symposium in Robot and Human Interactive Communication. IEEE, 2010, pp. 470–475.

[11] B. Görer, A. A. Salah, and H. L. Akın, "A robotic fitness coach for the elderly," in International Joint Conference on Ambient Intelligence. Springer, 2013, pp. 124–139.

[12] S. Jeong, C. Breazeal, D. Logan, and P. Weinstock, "Huggable: the impact of embodiment on promoting socio-emotional interactions for young pediatric inpatients," in Proceedings of the 2018 CHI Conference on Human Factors in Computing Systems, 2018, pp. 1–13.

[13] H. Javed, R. Burns, M. Jeon, A. M. Howard, and C. H. Park, "A robotic framework to facilitate sensory experiences for children with autism spectrum disorder: A preliminary study," ACM Transactions on Human-Robot Interaction (THRI), vol. 9, no. 1, pp. 1–26, 2019.

[14] T. Abbas, V.-J. Khan, U. Gadiraju, E. Barakova, and P. Markopoulos, "Crowd of oz: a crowd-powered social robotics system for stress management," Sensors, vol. 20, no. 2, p. 569, 2020.

[15] B. E. Wampold, "How important are the common factors in psychotherapy? an update," World Psychiatry, vol. 14, no. 3, pp. 270–277, 2015.

[16] J. Qinaau and A. Masuda, "Cultural considerations in the context of establishing rapport: A contextual behavioral view on common factors," in Handbook of Cultural Factors in Behavioral Health. Springer, 2020, pp. 75–92.

[17] G. M. Lucas, A. Rizzo, J. Gratch, S. Scherer, G. Stratou, J. Boberg, and L.-P. Morency, "Reporting mental health symptoms: breaking down barriers to care with virtual human interviewers," Frontiers in Robotics and AI, vol. 4, p. 51, 2017.

[18] L. D. Riek, P. C. Paul, and P. Robinson, "When my robot smiles at me: Enabling human-robot rapport via real-time head gesture mimicry," Journal on Multimodal User Interfaces, vol. 3, no. 1-2, pp. 99–108, 2010.

[19] T. W. Bickmore and R. W. Picard, "Establishing and maintaining long-term human-computer relationships," ACM Transactions on Computer-Human Interaction (TOCHI), vol. 12, no. 2, pp. 293–327, 2005.

[20] M. E. Seligman and M. Csikszentmihalyi, "Positive psychology: An introduction," in Flow and the foundations of positive psychology. Springer, 2014, pp. 279–298.

[21] M. E. Seligman, T. Rashid, and A. C. Parks, "Positive psychotherapy." American psychologist, vol. 61, no. 8, p. 774, 2006.

[22] S. M. Lamers, G. J. Westerhof, V. Kovács, and E. T. Bohlmeijer, "Differential relationships in the association of the big five personality traits with positive mental health and psychopathology," Journal of Research in Personality, vol. 46, no. 5, pp. 517–524, 2012.

[23] A. Villieux, L. Sovet, S.-C. Jung, and L. Guilbert, "Psychological flourishing: Validation of the french version of the flourishing scale and exploration of its relationships with personality traits," Personality and Individual Differences, vol. 88, pp. 1–5, 2016.

[24] P. L. Hill, N. A. Turiano, M. D. Hurd, D. K. Mroczek, and B. W. Roberts, "Conscientiousness and longevity: an examination of possible mediators." Health Psychology, vol. 30, no. 5, p. 536, 2011.

[25] C. E. Bartley and S. C. Roesch, "Coping with daily stress: The role of conscientiousness," Personality and individual differences, vol. 50, no. 1, pp. 79–83, 2011.

[26] K. Kokko, A. Tolvanen, and L. Pulkkinen, "Associations between personality traits and psychological well-being across time in middle adulthood," Journal of Research in Personality, vol. 47, no. 6, pp. 748–756, 2013.

[27] L. C. Quilty, F. De Fruyt, J.-P. Rolland, S. H. Kennedy, P. F. Rouillon, and R. M. Bagby, "Dimensional personality traits and treatment outcome in patients with major depressive disorder," Journal of Affective Disorders, vol. 108, no. 3, pp. 241–250, 2008.

[28] R. M. Niemiec, "Via character strengths: Research and practice (the first 10 years)," in Well-being and cultures. Springer, 2013, pp. 11–29.

[29] R. T. Proyer, F. Gander, S. Wellenzohn, and W. Ruch, "Strengths-based positive psychology interventions: A randomized placebo-controlled online trial on long-term effects for a signature strengths- vs. a lesser strengths-intervention," Frontiers in psychology, vol. 6, p. 456, 2015.

[30] R. T. Proyer, et. al, "Positive psychology interventions in people aged 50–79 years: long-term effects of placebo-controlled online interventions on well-being and depression," Aging & Mental Health, vol. 18, no. 8, pp. 997–1005, 2014.

[31] M. E. Seligman, T. A. Steen, N. Park, and C. Peterson, "Positive psychology progress: empirical validation of interventions." American psychologist, vol. 60, no. 5, p. 410, 2005.

[32] J. L. Smith and A. A. Hanni, "Effects of a savoring intervention on resilience and well-being of older adults," Journal of Applied Gerontology, vol. 38, no. 1, pp. 137–152, 2019.

[33] M. B. Donnellan, F. L. Oswald, B. M. Baird, and R. E. Lucas, "The mini-ipip scales: tiny-yet-effective measures of the big five factors of personality." Psychological assessment, vol. 18, no. 2, p. 192, 2006.

[34] E. Kallay and C. Rus, "Psychometric properties of the 44-item version of ryff's psychological well-being scale," European Journal of Psychological Assessment, 2014.

[35] J. D. Mayer and Y. N. Gaschke, "The experience and metaexperience of mood." Journal of personality and social psychology, vol. 55, no. 1, p. 102, 1988.

[36] M. Hesse, "The readiness ruler as a measure of readiness to change poly-drug use in drug abusers," Harm reduction journal, vol. 3, no. 1, p. 3, 2006.

[37] I.-U. C. for Political and S. Research., "National survey of midlife development in the united states (midus ii), 2004–2006: Documentation of psychosocial constructs and composite variables in midus ii project 1," 2010.

[38] T. Munder, F. Wilmers, R. Leonhart, H. W. Linster, and J. Barth, "Working alliance inventory-short revised (wai-sr): psychometric properties in outpatients and inpatients," Clinical Psychology & Psychotherapy: An International Journal of Theory & Practice, vol. 17, no. 3, pp. 231–239, 2010.

[39] V. Braun and V. Clarke, "Using thematic analysis in psychology," Qualitative research in psychology, vol. 3, no. 2, pp. 77–101, 2006.

[40] P. C. Heaven, K. Mulligan, R. Merrilees, T. Woods, and Y. Fairooz, "Neuroticism and conscientiousness as predictors of emotional, external, and restrained eating behaviors," International Journal of Eating Disorders, vol. 30, no. 2, pp. 161–166, 2001.

[41] K. A. Smith, M. G. Barstead, and K. H. Rubin, "Neuroticism and conscientiousness as moderators of the relation between social withdrawal and internalizing problems in adolescence," Journal of youth and adolescence, vol. 46, no. 4, pp. 772–786, 2017.

[42] B. B. Lahey, "Public health significance of neuroticism." American Psychologist, vol. 64, no. 4, p. 241, 2009.

[43] A. Sano, "Measuring college students' sleep, stress, mental health and wellbeing with wearable sensors and mobile phones," Ph.D. dissertation, Massachusetts Institute of Technology, 2016.